\documentclass[twocolumn,prl,amsmath,amssymb]{revtex4}
\usepackage{graphicx}
\usepackage{dcolumn}
\usepackage{bm}

\begin{document}
\title{Temperature dependent phonon renormalization in metallic nanotubes}
\author{V. Scardaci$^1$}
\author{S. Piscanec$^1$}
\author{Michele Lazzeri$^2$}
\author{R. Krupke$^3$}
\author{Francesco Mauri$^2$}
\email{mauri@lmcp.jussieu.fr}
\author{A.C. Ferrari$^1$}
\email{acf26@eng.cam.ac.uk}
 \affiliation{$^1$Cambridge University, Engineering Department, 9 JJ Thomson Avenue, Cambridge CB3 OFA, United Kingdom\\
$^2$Institut de Mineralogie et de Physique des Milieux Condenses,
Paris, France\\
$^3$Institut f\"ur Nanotechnologie Forschungszentrum Karlsruhe,
Germany }
\date{\today}

\begin{abstract}
We measure the temperature dependence of the Raman spectra of
metallic and semiconducting nanotubes. We show that the different trend in metallic tubes is due to
phonon re-normalization induced by the variation in electronic temperature, which is modeled including non-adiabatic
contributions to account for the dynamic, time dependent nature
of the phonons.
\end{abstract}

\maketitle

Raman spectroscopy is a powerful non destructive technique for the
characterization of carbon materials, and is a fundamental tool in
the recent advances in the understanding of single wall carbon
nanotubes (SWNTs). Raman experiments probe the optical
phonons, allowing to assess the vibrational properties of the
analyzed materials. A strong interplay exists between temperature
($T$) and phonon frequencies. Indeed, because of anharmonic
effects in the atomic oscillations, an increase in $T$ usually
results in a downshift of the phonon energy and a lifetime reduction~\cite{Ipatova1967}. For Raman active modes, this
corresponds to a downshift and a broadening of the Raman
peaks~\cite{Ipatova1967}.

The temperature dependence of the Raman spectra is extremely
effective for the evaluation of the local heating in a variety of
electronic devices~\cite{Jellison1983,Kuball2005}, and provides
valuable information for the characterization of
nano-materials~\cite{Piscanec2003}. Since SWNTs are at the center
of nanotechnology research, a thorough investigation and
understanding of the temperature effects on their Raman spectra is
needed, especially in view of their foreseen application in high current nanodevices. Several groups reported
changes of the Raman spectra of single-, double- and multi-wall
tubes as a function of T. Some focussed on the $G$
band~\cite{ Zhang2006, Zhou2006, Chiashi2005, Zhou2004,
Ouyang2004, Ci2003, Raravikar2002, Li2000, Huang1998}.  Others
considered the position~\cite{ Li2000, Raravikar2002, Fantini2004,
Zhou2004, Chiashi2005, Zhou2006, Zhang2006} and the
intensity~\cite{Fantini2004} of the Radial Breathing Modes (RBM).
A few reported the temperature evolution of the
2D~\cite{Huang1998, Zhang2006} and 2D' modes~\cite{Zhang2006}.

However, the different components of the $G$ band, which crucially
distinguish metallic from semiconducting nanotubes, were not
independently studied, in order to ascertain if those would have a
different temperature evolution in semiconducting and metallic
SWNTs, thus fingerprinting each material.

Here we present an extensive set of temperature-dependent
measurements of the $G^+$ and $G^-$ peaks in metallic
and semiconducting SWNTs.  We show that there is a significant
difference in the measured trends. We detect a
re-normalization of the $G^-$ peak, ruled by the variation in the electronic temperature, in metallic SWNTs. Furthermore, we prove that this
can only be explained by considering the dynamic nature of the
phonons and the resulting breakdown of the Born-Oppenheimer
approximation. The re-normalized phonon
frequencies in the standard static approaches being in total
disagreement with the experiments.

\begin{figure}
\includegraphics[width=80mm]{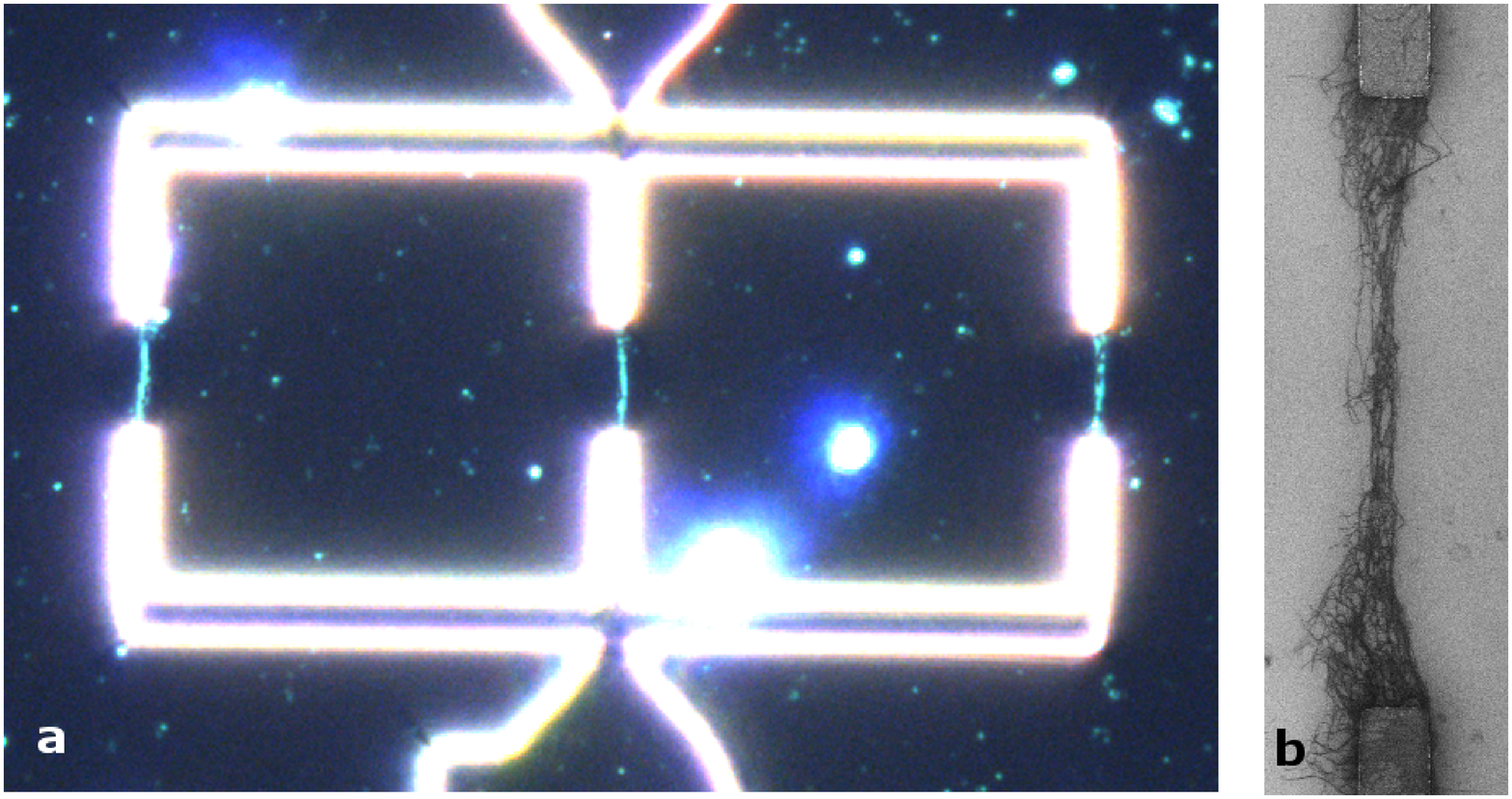}\\
\includegraphics[width=85mm]{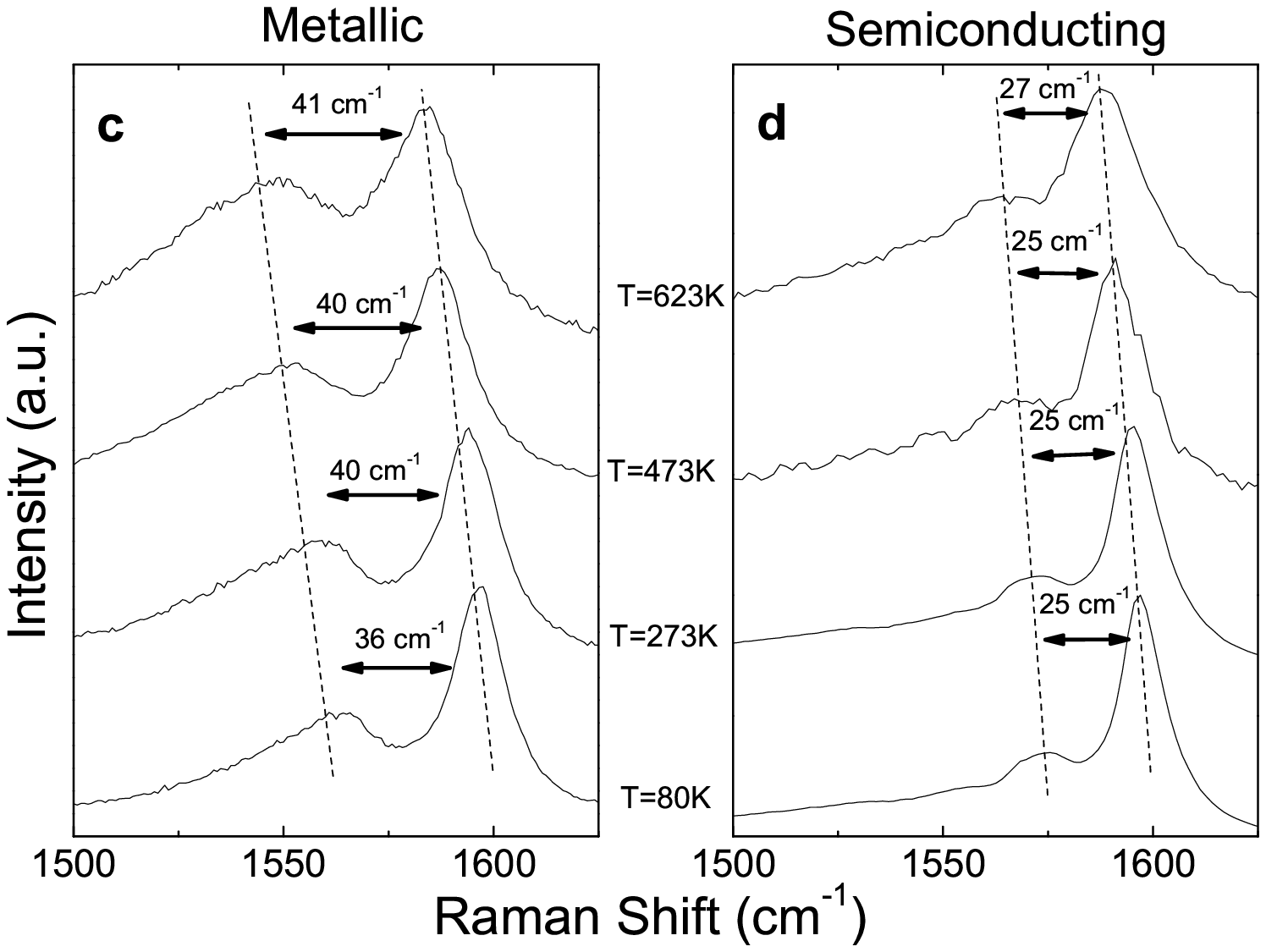}
\caption{(a) Dark-field microscopy image of the
electrodes used in the dielectrophoretic deposition, (b) SEM image
of the metallic tubes attached to the electrodes after
dielectrophoresis and Raman spectra of (c) metallic and (d)
semiconducting tubes acquired at T=80K, 273K, 473K and 623K.}
\label{Fig3}
\end{figure}

\begin{figure}
\includegraphics[width=62.5mm]{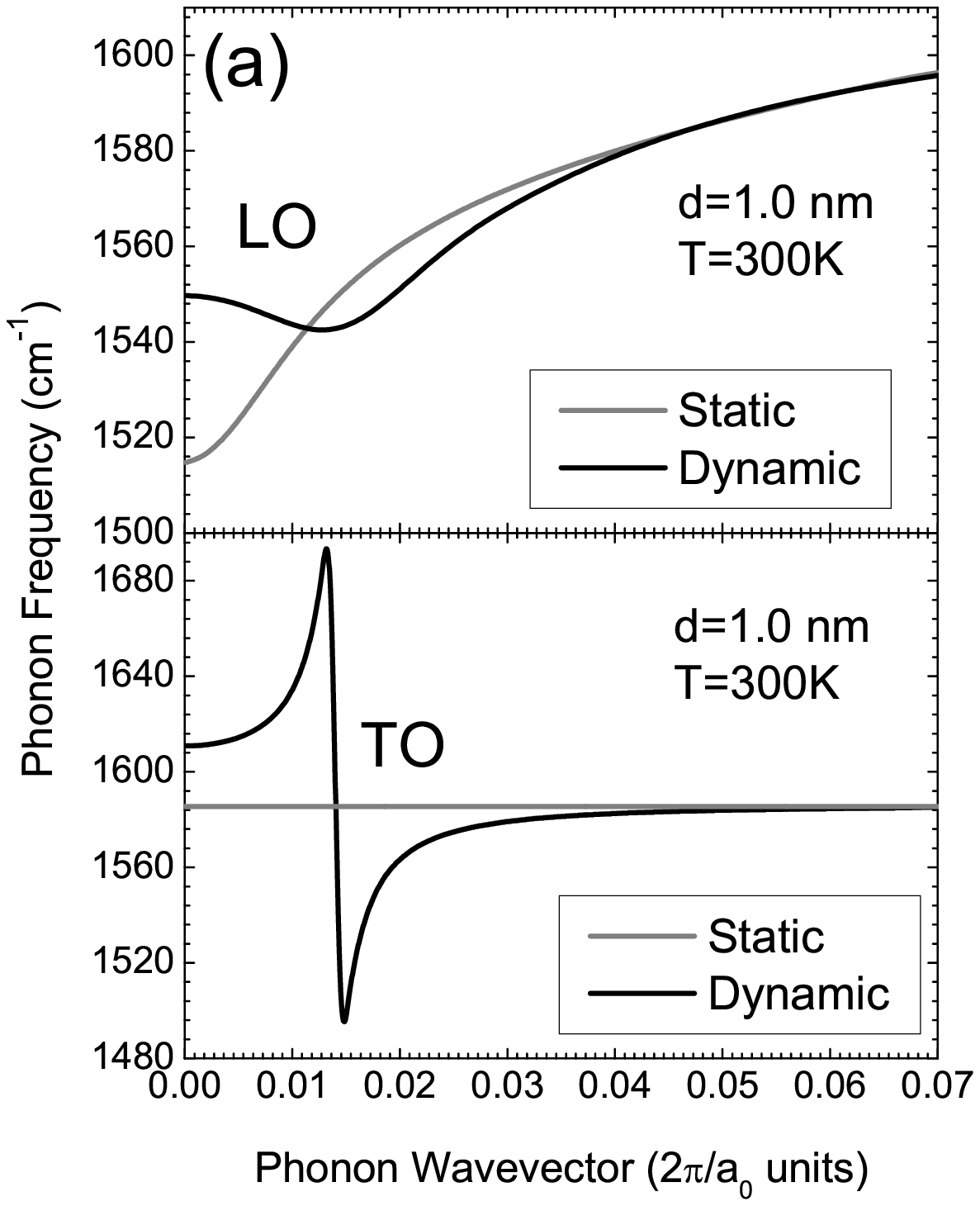}
\includegraphics[width=62.5mm]{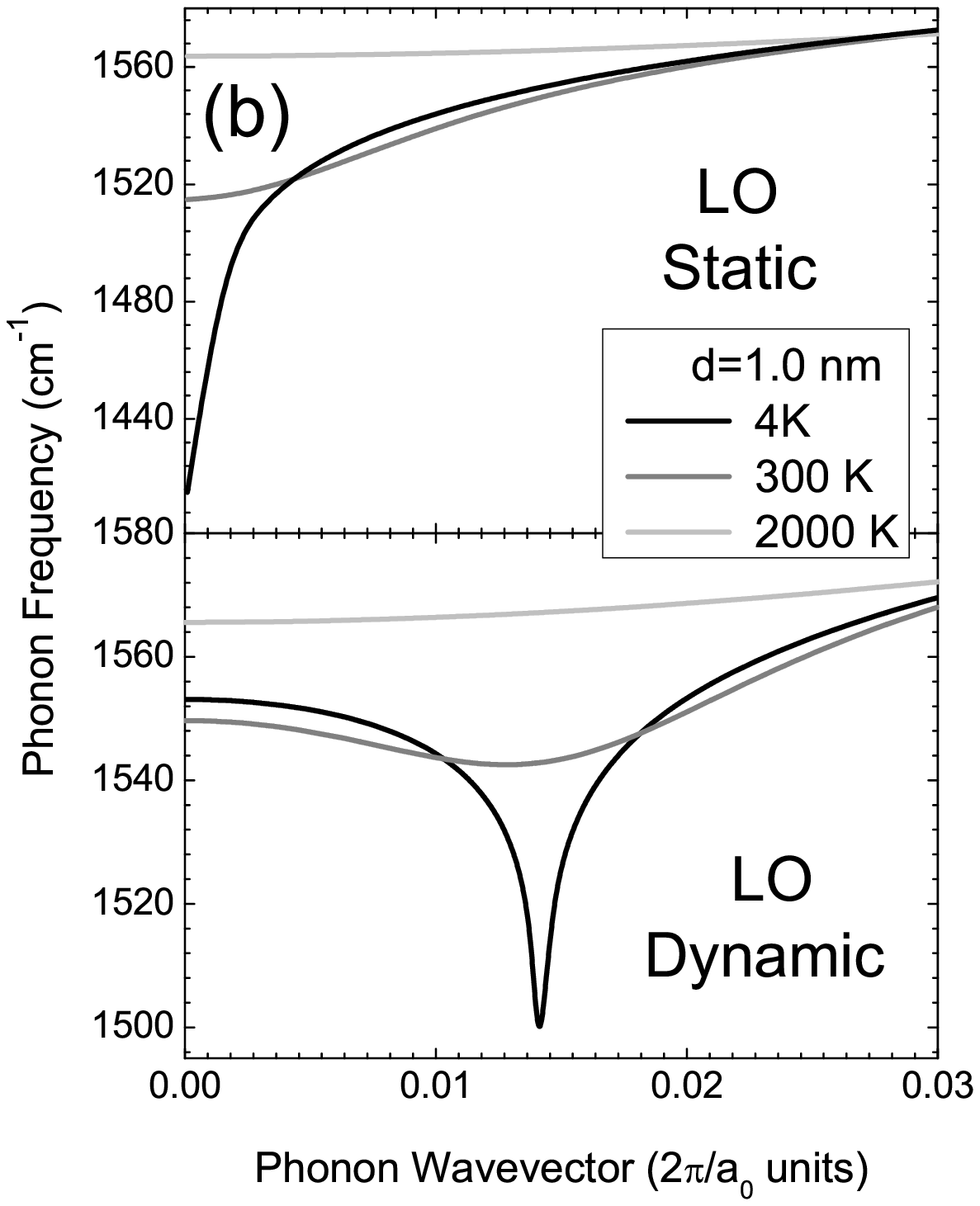}
\caption{(a)Comparison of the static and dynamic phonon dispersions of the LO and TO
modes close to q=0. (b) Static and dynamic
phonon dispersion of the LO mode close to q=0 calculated at
$T_e=$4K, 300K, and 2000K.} \label{Fig1}
\end{figure}

Samples containing almost exclusively metallic SWNTs are prepared
by depositing individually dispersed SWNT onto micro-electrodes
using radio frequency dielectrophoresis as described in
Ref.~\cite{Krupke2003, Krupke2004}. The SWNT suspension is
obtained by sonicating 100mL D$_2$O mixed with 0.05 weight \%
HiPCo-tube raw material provided by ``Houston'' and 1 weight \% of
sodium dodecylbenzene sulfonate (SDBS) (Sigma-Aldrich). The
suspension is centrifuged at 154000 g for 2 h and the upper 90\%
of the supernatant decanted. This is then diluted with D$_2$O to
obtain a surfactant concentration of 0.1 weight \% close to the
critical micelle concentration (0.097 weight \%). Microelectrodes
with 10 $\mu$m gap are then produced from gold by standard
electron beam lithography and bond-wired to a function generator.
Dielectrophoretic deposition is conducted by exposing the
electrodes driven at a frequency of 3 MHz and a peak-to-peak
voltage V$_{pp}$ of 20 V to a drop of suspension ($\sim 10 \mu$L).
After 10 min the samples are subsequently rinsed with H$_2$O,
ethanol and dried in nitrogen. They are then characterized by
optical dark-field microscopy (Fig.~\ref{Fig1}a), electron
microscopy (Fig.~\ref{Fig1}b) and Raman spectroscopy
(Fig.~\ref{Fig1}c), which show that aligned bundles of metallic
SWNTs bridge the electrodes.

Temperature dependent Raman measurements are carried out with a
Renishaw 1000 spectrometer, using a He-Ne laser at 633 nm, matching the resonance window for HiPCo metallic tubes~\cite{
Kataura1999}. To further enhance the signal, the laser
is polarized parallel to the aligned tubes. The temperature of the
sample is set by a Linkam stage, in the range 77-850K. The
stage is cooled by liquid N$_2$. Raman spectra are recorded
from 80 to 630 K with steps of 25 K. From the RBM frequency
~\cite{Kataura1999, Jorio2001, Telg2004} we estimate the mean
diameter of metallic tubes to be $\sim 1.0$ nm.

We also acquire spectra from samples with a natural mixture
of metallic and semiconducting HiPCo-tubes. There we use 514 nm
excitation, which brings
semiconducting SWNTs with a diameter of $\sim 1.1$ nm in resonance, as
confirmed by the analysis of the RBMs. The semiconducting nature
of the measured tubes is also confirmed by the shape and position of the $G^-$-peak~\cite{Piscanec2007}.

The $G^+$ and $G^-$ peaks of metallic and semiconducting tubes
measured at 80 K, 273 K, 473 K, and 625K are shown in
Fig.~\ref{Fig3}. To quantitatively determine the peak positions, we fit the two components of the $G$
band using 2 Lorentzians. An increasing temperature results in a
downshift of both $G^+$ and $G^-$. This is in
agreement with all previous observations~\cite{Zhang2006,
Zhou2006, Chiashi2005, Zhou2004, Ouyang2004, Ci2003,
Raravikar2002, Li2000, Huang1998} and can be explained by
an-harmonicity. However, we crucially detect that in
semiconducting tubes the splitting between $G^+$ and $G^-$ is
$\sim25$ cm$^{-1}$, independent of temperature, while in metallic it increases with $T$. This
different behavior seems unlikely to originate from
anharmonicity.

The $G^+$ and $G^-$ peaks originate from the tangential (TO) and
the longitudinal (LO) modes derived from the splitting of the
$E_{2g}$ phonon of graphene. In metallic tubes, the LO mode is
affected by a Kohn anomaly (KA)~\cite{Kohn1959}, which causes the
softening of this phonon~\cite{Lazzeri2006, Piscanec2007}. Since
KA are not present in semiconducting SWNTs, the $G^+$, $G^-$
assignment in metallic SWNTs is the opposite of
semiconducting tubes~\cite{Lazzeri2006, Piscanec2007}.

For a given temperature and SWNT diameter, the softening of the LO
mode in metallic with respect to semiconducting SWNTs originates
from an anomalous screening of the atomic displacements, due to
the electrons close to the Fermi energy. The occupation of these
electronic states depends on temperature through the Fermi-Dirac
distribution. Thus, in principle, changes in temperature may
result in a modification of the electronic screening (see, for
example, Ref.~\cite{ZimanBook}), which can re-normalize the LO
phonon frequency and the position of its corresponding Raman peak.
This contribution, if present, would depend only on the electronic
temperature $T_e$, and has to be added to the an-harmonic effects.

The influence of $T_e$ on the screening of the atomic vibrations
can be investigated theoretically. In general, lattice
dynamics is calculated within the Born-Oppenheimer adiabatic
framework~\cite{Born1927, BornBook}. This assumes the motions of
ions and electrons to be completely decoupled, with the electrons
following adiabatically the ions (in other words, the electrons
always ``see" the ions as if they are in fixed positions). This is justified when the occupied and empty states are separated by an energy gap~\cite{MaradudinBook}. In
materials without an electronic gap, the Born-Oppenheimer approximation (BOA) is not
easily justifiable, however experience proves that in most
cases this accurately reproduces the phonon dispersion of
metals~\cite{Degironcoli1995}.

Phonon calculations within the BOA and based on zone-folding are in good agreement with room T Raman measurements of SWNTs~\cite{Lazzeri2006}.
However, we have shown that the BOA breaks
down in the description of KA in doped
graphene~\cite{Lazzeri2006graph,Pisana2007}. Therefore, it is necessary to probe whether an adiabatic approach can
reproduce the T evolution of the SWNTs Raman peaks.

Phonons can be regarded as a perturbation of a
crystal~\cite{Eschrig1973}. In general, given their dynamic
nature, they should be described by time-dependent
perturbation theory (TDPT)~\cite{Eschrig1973}. However, within the
adiabatic BOA, they are seen as
static perturbations and treated by a time-independent
perturbation theory (TIPT)~\cite{Eschrig1973}.

Starting from the general expression given in
Ref.~\cite{Allen1980} and following the approach described in
Ref.~\cite{Pisana2007}, it can be shown that, within TDPT, the
reciprocal-space expression for the non analytic part of the
dynamical matrix of SWNTs, $\Theta_{q}$ is given by:
\begin{eqnarray}
\tilde{\Theta}_q&=&\frac{4\tau A_{\Gamma/K}}{2\pi}
\sum_{m,n=L,R}\int_{-\bar
k}^{\bar k} |D_{(K+k'+q)n,(K+k')m}|^2 \nonumber \\
&& \frac{f_{K+k,m}-f_{K+k'+q,n}}
{\epsilon_{K+k',m}-\epsilon_{K+k'+q,n}+\hbar\omega_{q}+i\gamma}
dk',
 \label{Dtilde dynamic}
\end{eqnarray}

where $\tau$ is the length of the translational unit cell of the tube,
$k'$ is measured from the Fermi point $k_{F}$; $\bar k$ has a
small but finite value; R and L label the bands crossing at the
Fermi energy and corresponding respectively to right- and
left-moving electrons; $A_{\Gamma/\rm K}$ accounts for the number
of processes satisfying 2$q=k_F$ ($A_{\Gamma}=2$,
$A_{K}=1$),$\hbar\omega_{q}$ is the energy of a phonon of
wavevector $4$ and branch $\eta$ (omitted in the equations for
simplicity), $\gamma$ is a small real number, $D_{(k+q)i,kj} =
\langle k+q,i| \Delta V_{q} |k,j\rangle$ is the electron-phonon
coupling (EPC) matrix element where $|k,i\rangle$ is the
electronic Bloch eigenstate of wavevector $k$, band $i$, energy
$\epsilon_{k,i}$, and occupation $f_{k,i}$ given by the
Fermi-Dirac distribution~\cite{AshcroftBook}; $\Delta
V_{q}$ is the derivative of the electronic potential with respect
to the phonon normal coordinate.

Within the BOA, a phonon is seen in its
static limit, i.e. assuming $\omega_{q}=0$ and $\gamma=0$. $\tilde{\Theta}_{q}$ thus
simplifies in:
\begin{eqnarray}
\tilde{\Theta}_{{\rm q}}&=&\frac{2A_{\Gamma/\rm
K}T}{2\pi}\sum_{m,n=L,R}\int_{-\bar k}^{\bar k} \frac{f_{{\rm
{K+k'}},m}-f_{{\rm K} + {\rm k'} + {\rm q},n}} {\epsilon_{{\rm
{K+k'}},m}-\epsilon_{{\rm K}+{\rm k'}+{\rm q},n}} \nonumber \\
&& |D_{({\rm K}+{\rm k}+{\rm q})n,({\rm K}+{\rm k'})m}|^2. dk'.
\label{Dtilde}
\end{eqnarray}
KAs occur for phonons (i) having non-zero EPC between states close
to the Fermi energy, and (ii) for which the denominators in
Eq.~\ref{Dtilde dynamic},\ref{Dtilde} vanish, resulting in the
presence of a singularity in the dynamical
matrix~\cite{Piscanec2004}. Thus, within a static approach, the
anomalies are predicted to occur for the values of $q$ that make
the denominator in Eq.~\ref{Dtilde} vanish, i.e. for $q=0$ and
$q=2k_F$~\cite{Piscanec2004, Lazzeri2006, Dubay2002, Bohnen2004,
Connetable2005}.

Thus, TDPT deeply modifies the
description of the KAs. Because of the $\hbar
\omega_q + i\gamma$ terms, and assuming the electronic bands of
the SWNTs at the Fermi energy to be linear with slope $\beta$, the
denominator of Eq.~\ref{Dtilde dynamic} vanishes for $q=\pm\hbar
\omega_q/\beta$ and $q=k_F\pm\hbar \omega_q/\beta$, resulting in a
shift of the position of the KAs. We refer to the inclusion of
these terms as the \emph{dynamic effects}.

By using the folding approach described in
Refs.~\cite{Lazzeri2006, Piscanec2007}, we numerically integrate
Eq.~\ref{Dtilde dynamic},~\ref{Dtilde}, and obtain the theoretical
description of the KAs in metallic SWNTs within the static and the
dynamic approaches. These results are then corrected to account for curvature~\cite{Piscanec2007}.

Fig.~\ref{Fig1} shows the phonon dispersion of the LO and TO
modes close to $q=0$ for a metallic SWNT with $d=1.0$ nm.
Calculations are done within the static and the dynamic
approaches, at room T. The first predicts a KA
for the LO mode only. This is centered at $q=0$, and the
phonon dispersion close to it has a logarithmic shape. On the
other hand, dynamic calculations show that both the LO and TO
phonon branches are affected by KA, and that the anomalies are
centered at $q=\hbar\omega_q/\beta$.

Eq.~\ref{Dtilde dynamic},\ref{Dtilde} depend on the occupation of
the electronic states through the Fermi-Dirac distribution. Thus,
it is possible to compute the dependence of the phonon frequencies
on the electronic temperature.

To stress that these effects depend uniquely on the occupation of
the electronic bands of the tubes, here we distinguish between the
electronic and the ionic temperature, which we indicate with $T_e$
and $T_i$ respectively.
$T_i$ corresponds to the energy associated to the atoms vibrating
around their equilibrium positions, and determines the onset of
the an-harmonic effects, while $T_e$ fixes the electronic states
population according to the Fermi-Dirac distribution and
determines the shape of the KA. At thermal equilibrium $T_i=T_e=T$.

Fig.~\ref{Fig1} shows the KA of the LO branch for a metallic SWNT
of $d=1.0$ nm calculated at $T_e$=4K, 300K and 2000K within the
static and the dynamic frameworks. In both cases, increasing $T_e$
results in a softening of the anomaly, this being
stronger if the dynamic effects are taken into account. The TO branches do not show any dependence on $T_e$ in
both the static and the dynamic approach. Thus, the static and the dynamic models
predict a totally different temperature dependence of the
splitting between the LO and the TO phonons at $q=0$.

The LO and TO frequencies are also modified by the presence of
an-harmonic effects, which are related to
$T_i$ and are expected to give an overall decrease of the
phonon frequencies for increasing T. These are not
described by Eq.~\ref{Dtilde dynamic},\ref{Dtilde}. However, a
simple model for the T dependence of the frequency shift
for the LO and TO modes predicts them to be the
same~\cite{Grimvall}. Thus, the \textit{relative} position of these two
modes is determined \textit{only} by the dependence on $T_e$,
which is entirely described by Eq.~\ref{Dtilde
dynamic},\ref{Dtilde}. Furthermore, since KAs affect metallic SWNTs only, no dependence
of the LO-TO splitting on $T_e$ should be observed for
semiconducting SWNTs.

\begin{figure}
\centerline{\includegraphics[width=75mm]{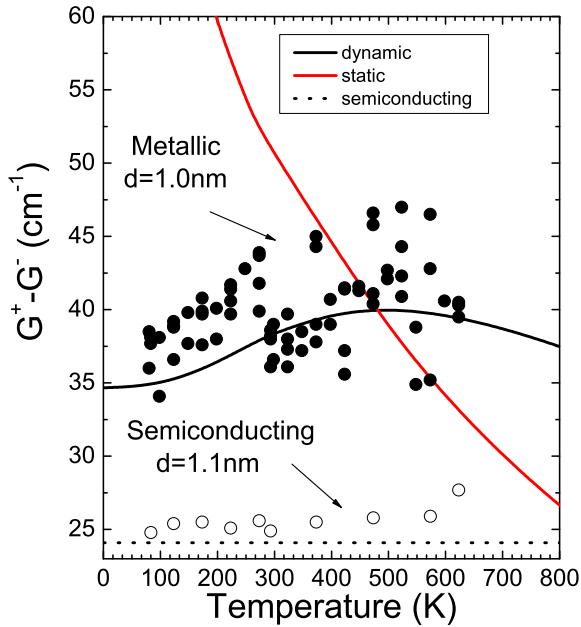}}
\caption{Temperature dependence of the G$^+$ and G$^-$ splitting
compared with the theoretical prediction of the LO-TO splitting
obtained using static and dynamic approaches.} \label{Fig4}
\end{figure}

The experimental LO-TO splitting and those calculated by using the
static and dynamic theory are plotted in Fig.~\ref{Fig4}. The
static and the dynamic models give contrasting predictions. In the
static case, the splitting for metallic SWNTs with $d=1.0$ nm is
estimated to be 80 cm$^{-1}$ at 80 K, and to decrease to 50
cm$^{-1}$ and 30 cm$^{-1}$ at 300 K and 625 K respectively. On the
other hand, the dynamic model predicts an increase of the
splitting from 35 cm$^{-1}$ to 40 cm$^{-1}$ when moving from 80 to 520 K, and a plateau in the 420-650 K region. Both models
predict a reduction of the splitting for T$>$650 K, with a steeper slope for the static case. For semiconducting tubes, no differences exist
between the static and dynamic predictions, and the splitting for $d=1.1$ nm is 24 cm$^{-1}$.

The static model trend is clearly incompatible with the
experiments, which, on the opposite, are in very good
agreement with the predictions of the dynamic approach.
Indeed, the experimental splitting for metallic tubes shows a
slight increase with temperature and peaks at $\sim$45 cm$^{-1}$ .
On the other hand, no temperature dependence is observed for
semiconducting tubes, for which the splitting is $\sim$25cm$^{-1}$.

In conclusion, we presented the temperature dependence of the
$G^+$ and $G^-$ peaks in the Raman spectra of SWNTs. We showed that the temperature dependence of
the $G^+$ and $G^-$ splitting is significantly different in
metallic and semiconducting SWNTs. The increase of the $G^+$ and
$G^-$ splitting in metallic tubes is due to a
non-adiabatic phonon re-normalization induced by the variation in
electronic temperature. This cannot be described by the
standard adiabatic Born-Oppenheimer approach.

S.P. acknowledges funding from the Maudslay Society and from
Pembroke College, A. C. F. from EPSRC grants GR/S97613,
EP/E500935/1 the Royal Society and the Leverhulme Trust.


\begin{thebibliography}{100}

\bibitem{Ipatova1967}
I.~P. Ipatova, A.~A. Maradudin, and R.~F. Wallis,  Phys. Rev.
\textbf{155}, 882 (1967).

\bibitem{Jellison1983}
G.E. Jellison,F.A. Modine, Phys. Rev. B \textbf{27}, 7466
(1983).

\bibitem{Kuball2005}
M. Kuball et al., Phys. Status Solidi A \textbf{ 202}, 824 (2005).

\bibitem{Piscanec2003}
S. Piscanec et al., Phys. Rev. B \textbf{68}, 241312 (2003).

\bibitem{Zhang2006}
Q. Zhang et al., Smart Mat. and Struct. \textbf{15}, S1 (2006).

\bibitem{Zhou2006}
S.~Y. Zhou et al., Nature physics \textbf{2}, 595 (2006).

\bibitem{Chiashi2005}
S. Chiashi et al., Therm. Sci. Eng. \textbf{13}, 71 (2005).

\bibitem{Zhou2004}
Z.~P. Zhou et al., Chem. Phys. Lett. \textbf{396}, 372 (2004).

\bibitem{Ouyang2004}
Y. Ouyang and Y. Fang, Physica E \textbf{ 24}, 222 (2004).

\bibitem{Ci2003}
L. Ci et al., Appl. Phys. Lett. \textbf{82}, 3098 (2003).

\bibitem{Raravikar2002}
N.~R. Raravikar et al., Phys. Rev. B \textbf{66}, 235424 (2002).

\bibitem{Li2000}
H.~D. Li et al., Appl. Phys. Lett. \textbf{76}, 2053 (2000).

\bibitem{Huang1998}
F.~M. Huang et al., J. Appl. Phys. \textbf{84}, 4022 (1998).

\bibitem{Fantini2004}
C. Fantini et al., Phys. Rev. Lett. \textbf{93}, 147406 (2004).

\bibitem{Krupke2003}
R. Krupke et al., Science, \textbf{301}, 344 (2003).

\bibitem{Krupke2004} R. Krupke et al., Nano Lett.
\textbf{4}, 1395 (2004).

\bibitem{Kataura1999}
H. Kataura et al., Synth. Met. \textbf{103}, 2555 (1999).

\bibitem{Telg2004}
H. Telg et al., Phys. Rev. Lett. \textbf{93}, 177401 (2004).

\bibitem{Jorio2001}
A. Jorio et al., Phys. Rev. Lett. \textbf{86},  1118 (2001).

\bibitem{Piscanec2007}
S. Piscanec et al., Phys. Rev. B \textbf{75} 035427 (2007).

\bibitem{Kohn1959}
W. Kohn, Phys. Rev. \textbf{2}, 393 (1959).

\bibitem{Lazzeri2006}
M. Lazzeri et al., Phys. Rev. B \textbf{73}, 155426 (2006).

\bibitem{ZimanBook}
J.~M. Ziman, \emph{Principles of the theory of solids} (Cambridge
University Press, London, New York, 1964).

\bibitem{Born1927}
M. Born, R. Oppenheimer, Ann. Phys. \textbf{84}, 457 (1927).

\bibitem{BornBook}
M. Born and K. Huang, \emph{Dynamical Theory of Crystal Lattices}
(Clarendon Press, Oxford, 1954).

\bibitem{MaradudinBook}
G.~K. Horton and A.~A. Maradudin, \emph{Dynamical Properties of
Solids Vol. 1} (North-Holland, Amsterdam, 1974).

\bibitem{Degironcoli1995}
S. Degironcoli, Phys. Rev. B \textbf{51}, 6773 (1995).

\bibitem{Pisana2007}
S. Pisana et al., Nature Mat. \textbf{ 6}, 198 (2007).

\bibitem{Lazzeri2006graph}
M. Lazzeri,F. Mauri, Phys. Rev. Lett.\textbf{97},266407 (2006).

\bibitem{Eschrig1973}
H. Eschrig, Phys. Stat. Sol. B \textbf{ 56}, 197 (1973)

\bibitem{Allen1980}
P.~B. Allen, \emph{Dynamical Properties of Solids Vol. 1}
(North-Holland, Amsterdam, 1980).

\bibitem{AshcroftBook}
N.~W. Ashcroft and N.~D. Mermin, \emph{Solid state physics}
(Saunders College, New York,London, 1976).

\bibitem{Piscanec2004}
S. Piscanec et al., Phys. Rev. Lett. \textbf{93}, 185503 (2004).

\bibitem{Dubay2002}
O. Dubay et al. Phys. Rev. Lett.
\textbf{88}, 235506 (2002).

\bibitem{Bohnen2004}
K.~P. Bohnen et al., Phys. Rev. Lett. \textbf{93}, 245501 (2004).

\bibitem{Connetable2005}
D. Connetable et al., Phys. Rev. Lett. \textbf{94}, 015503 (2005).

\bibitem{Grimvall}
G. Grimvall, \emph{Thermophysical properties of materials} (North
Holland, 1986)

\end{thebibliography}
\end{document}